\documentclass[reprint,amsmath,amssymb,aps]{revtex4-2}

\usepackage{graphicx}
\usepackage{dcolumn}
\usepackage{bm}
\usepackage{braket}

\begin{document}

\title{The validity of the universal relation between corrections to entropy
	and the extremality of Schwarzschild-de Sitter black holes under the GUP and EUP}

\author{Hongbo Cheng}
\email{hbcheng@ecust.edu.cn}
\author{Yinan Zhao}
\affiliation{
Department of Physics, East China University of Science and Technology, Shanghai 200237, China\\
The Shanghia Key Laboratory of Astrophysics, Shanghai 200234, China}

\date{\today}

\begin{abstract}
We investigate the extremality relations by examining perturbative corrections to both the entropy of Schwarzschild-de Sitter black holes and their extremality bounds under the Nariai limit within the frame of the generalized uncertainty principle (GUP) and the extended uncertainty principle (EUP) respectively. The GUP-corrected or EUP-corrected horizons bring about the additional terms finally in the Goon-Penco relation to violate the universal ones and these extra terms cannot be canceled by adding the adequate conditions. We argue that the corrected uncertainty principles including GUP and EUP violate the validity of extremality relations because no matching condition can be imposed to support the Goon-Penco relation unless the influences from GUP and EUP disappear. 
\end{abstract}

\maketitle


\section{Introduction}

The black holes have been attracting a lot of attentions from the physical community [1, 2]. The black holes can be thought as one kind of compact objects in the general relativity and are the final stage of the collapse of gravitational sources [1, 2]. Within the frame of classical gravity, the event horizon of black hole has not any outgoing geodesics but ingoing ones, which means that no matter leave and no radiation can escape from the central sources [1-3]. Classically we cannot receive any signals from this kind of sources. Further the black holes need to be explored in various directions. The black hole entropy relating to the surface area of the black hole was introduced [4, 5], so were the relevant variables such as internal energy, temperature etc. [6-8]. The four laws of black hole thermodynamics were certainly found [8]. Hawking perceived the black holes as thermal bodies radiating energy with the Hawking temperature proportional to surface gravity [6, 7]. Afterwards a lot of researches have been performed on the thermodynamic quantities and the so-called four thermodynamic laws which describe the black holes behaviours. In the context of black hole thermodynamics, there are more equations and relationships need to be considered and solved. In the quantum gravity theory, the Weak Gravity Conjecture (WGC) as a criteria was proposed [9]. The WGC implies that the gravity is the weakest force in a system with mass $M$ and charge $Q$, expressed as $\frac{Q}{M}>1$, but $\frac{Q}{M}\leq 1$ for the extremal black holes [10]. In order to hold the WGC for the extremal inversing the charge to mass ratio, the higher-derivative operators with their own coefficients were added and the coefficients should be adjusted, which leading the Black Hole Weak Gravity Conjecture [11]. The generalized WGC involving the higher-derivative corrections can also be obeyed for the evaporating black holes emitting the uncharged particles [12, 13]. The WGC can bring about the correction terms to rule out the naked singularities [14]. In other words, the removal of naked singularities implies the validity of the WGC inequality [14]. Goon and Penco researched on the universality of the thermodynamic relation between the entropy and extremality of black holes under the perturbation in order to further the investigations on the WGC [15]. The Goon-Penco relation shows the WGC-like behaviour of black hole in the form of a universal thermodynamic extremality relation [15],
\begin{align}
	\frac{\partial M_{ext}(\overrightarrow{Q},\varepsilon)}{\partial\varepsilon}=\lim_{M\rightarrow M_{ext}(\overrightarrow{Q},\varepsilon)}[-T(\frac{\partial S(M,\overrightarrow{Q},\varepsilon)}{\partial\varepsilon})_{M,\overrightarrow{Q}}],
\end{align}
where the extremality bound $M_{ext}(\overrightarrow{Q}, \varepsilon)$ satisfies $M_{ext}(\overrightarrow{Q}, \varepsilon)<M$, $M$ as mass of black hole. Here $\overrightarrow{Q}$ has several components standing for the thermodynamic quantities. The parameter $\varepsilon$ provides some insights into the extremality bound, entropy and the equation (1). $T=T(M, \overrightarrow{Q}, \varepsilon)$ is the temperature while a function of variables. The Eq.(1) is a universe relation connecting the derivatives of mass-like function with the entropy function with respect to the perturbation $\varepsilon$. The relation declares that with respect to the control parameters of correction, the black hole entropy increases while the mass of extremal black hole decreases. The contributions of Ref.[14] also supported the WGC in the flat spacetime. Some works have been finished for various charged and rotating black holes in the AdS spacetimes and the Goon-Penco relation were confirmed repeatedly [16, 17]. In addition the universal relation by Goon and Penco between the shifted entropy and mass bound was checked and generalized for the rotating AdS black holes [18]. There exist the universal relations for the charged-rotating-AdS black holes surrounded by the quintessence with the cloud of strings and this kind of black holes have the WGC-like behaviour [19]. The universal relations for the black holes in the de Sitter background such as Schwarzschild-de Sitter, Reissner-Nordstrom-de Sitter and Kerr-de Sitter black holes are considered and the validity of the relation can be kept with the corrections to the black holes [20]. The Goon-Penco universal relation is valid for accelerating black holes in anti-de Sitter spacetime with an appropriate matching condition which is needed between the perturbation parameter of the relation with the perturbative correction to the black holes and the parameter with information about the conical deficits on the north and south poles with the cosmic string tensions [21].

In the process of research on the quantum properties of black holes, we can not neglect the gravitational effect. The quantum characteristics of black hole such as metric fluctuations [22-24] and the quantum structures [25-27] have something to do with the black holes horizons. There are quantum gravity effects not to be ignored on the evaporation profile and singularity removal of microscopic black holes [28]. The Heisenberg uncertainty relation associated with the quantum effects must be altered under the gravitational influence and the terms involving the Newtonian constant will be added in the uncertainty principle [29-35]. The new Heisenberg uncertainty principle with the additional terms can be utilized to overcome the divergence from state density near the black hole horizon and relate the black hole entropy to a minimum length as quantum gravity scale [36-39]. Under the gravitational influence, several corrected forms of the Heisenberg uncertainty principle can be selected with terms as functions of the momentum difference or distance interval respectively [30, 40]. The inequality consisting of series terms based on the momentum difference can be thought as the generalized uncertainty principle (GUP) [30, 40]. The position-uncertainty-corrected Heisenberg inequality is perceived as the extended uncertainty principle (EUP) reflecting the quantum effects at an extremely large scale [30, 41-43]. The bounds on the parameters appearing in the GUP were estimated from gravitational wave data of the events GW150914 and GW190521 [44]. In view of the GUP and EUP, more researches proceeded. In the higher-dimensional spacetime, the relation between the Hawking tunneling radiation and a minimal length as the quantum gravity scale for the black holes were discussed in the context of the GUP [36, 37]. We computed the Parikh-Kraus-Wilczeck tunneling radiation of black holes with global $f(R)$ monopole to show that the square of the momentum difference term from the GUP promotes the emission [45]. We also indicated that the GUP effect leads the fragmentation instability of this kind of black holes [46]. The authors of Ref. [47] studied the charged particle absorption to exhibit the GUP influences on the thermodynamics and weak cosmic censorship conjecture in AdS black holes. It was found that the GUP can affect the thermodynamic effects of Bardeen black holes surrounded by perfect fluid dark matter [48]. Some problems in acoustic black holes under GUP were discussed [49]. The EUP can amend the innermost stable circular orbits around the black holes and the size of photosphere while the EUP contributes its correction to fit the Milky Way's rotation curve [28]. There are some contributions from EUP on the thermodynamic characteristics and Unruh effect of the Schwarzschild black hole [50]. It was shown that the influence of EUP causes the black holes swallowing the $f(R)$ global monopole not to be stable in the processes of Parikh-Kraus-Wilczeck tunneling radiation and the compact body's division [51].

It is significant to revisit the Goon-Penco relation for the Schwarzschild black holes in the universe with the cosmological radius subject to the corrected Heisenberg uncertainty listed as GUP or EUP. It is also significant to promote the research on the various black holes in the de Sitter spacetime under the Nariai limit although Ko and Gwak confirmed the validity of the universal relation [20]. We should explore the influence from GUP or EUP on the Goon-Penco relation for dS black holes with respect to the extremal condition from Ref. [20]. According to the previous works mentioned above, the corrections to the standard uncertainty principle must change the black hole horizons to have the extremality bound and the entropy rewritten undoubtedly and the original Goon-Penco relation is altered further. Maybe the special mass and the entropy of black hole do not vary synchronously under the influence from GUP or EUP, so we must revisit the universal relation for the Schwarzschild-de Sitter black holes while the inequality contains the Newton-constant-dependent terms. As a starting point, we focus our attention on the thermodynamic relation for Schwarzschild - de Sitter black holes in the Nariai spacetimes while the corrected Heisenberg uncertainties are introduced. We wonder how the GUP or EUP modifies the Goon-Penco universal relation and whether there exist the suitable matching conditions to hold the relation. According to our results we can further our exploration for the relations for Reissner-Nordstrom-de Sitter and Kerr-de Sitter black holes within the frame of Nariai issue under the GUP or EUP.

This paper is set up as follows. First we give a brief exposition of the works in Ref. [20] on the thermodynamics of Schwarzschild-de Sitter black holes and the thermodynamic variables of this kind of black holes under Nariai limit. Secondly we introduce the GUP and EUP and derive the Hawking temperature, entropy under the influence of gravitation uncertainty principles. We scrutinize the variations of the corrected mass and corrected entropy with respect to the perturbative parameter to test the universal relation in favour of the coincidence of the outer horizon and cosmological horizon. Finally we keep the differential equation to find the necessary conditions and put forward our arguments.

\section{The universal relation on the Schwarzschild-de Sitter black holes}

The purpose of this section is to review the researches in the universal relation for the Schwarzschild-de Sitter black holes from Ref. [20]. The Schwarzschild-de Sitter black holes are spherically symmetric solutions to the Einstein's equations with the cosmologicl constant, so the spacetime metric can be denoted as [1, 2, 20],
\begin{align}
ds^{2}=f(r)dt^{2}-\frac{dr^{2}}{f(r)}-r^{2}(d\theta^{2}+\sin^{2}\theta d\varphi^{2}).
\end{align}
According to Ref. [15], the action with the cosmological constant viewed as a perturbation parameter is [20],
\begin{align}
I=\frac{1}{16\pi}\int d^{4}x\sqrt{-g}(R-2(1+\varepsilon)\Lambda),
\end{align}
where $g$ is the determinant of the metric (2) and $R$ is the Ricci scalar. Here $\Lambda$ is the cosmological constant and the perturbative parameter $\varepsilon$ is tiny. Themetric function is a solution to the field equation corresponding to the action (3) and is obtained [20],
\begin{align}
	f(r)
	&=1-\frac{2M}{r}-\frac{1+\varepsilon}{3}\Lambda r^{2}\notag\\
&=1-\frac{2M}{r}-\frac{1+\varepsilon}{l^{2}}r^{2}
\end{align}
with mass of black hole $M$. Here the cosmological radius $l$ satisfies $\Lambda=\frac{3}{l^{2}}$. The entropy of black hole is [4, 5]
\begin{align}
	S=\frac{A_{H}}{4}=\pi r_{H}^{2},
\end{align}
where $A_{H}$ is the surface area of black hole. As a root of $f(r_{H})=0$, $r_{H}$ is the event horizon. We can set $f(r)=0$ and combine with the entropy (5) to obtain the shifted mass [20],
\begin{align}
	M=\frac{\sqrt{S}}{2\sqrt{\pi}}-\frac{1+\varepsilon}{2l^{2}}\frac{S^{\frac{3}{2}}}{\pi^{\frac{3}{2}}}.
\end{align}
The perturbative factor can be expressed in terms of $M$ and $S$ from Eq. (6)[20],
\begin{align}
	\varepsilon=-\frac{2l^{2}\pi^{\frac{3}{2}}}{S^{\frac{3}{2}}}M+\frac{\pi l^{2}}{S}-1.
\end{align}
The partial derivative to the parameter (7) with a fixed mass is made [20],
\begin{align}
	(\frac{\partial\varepsilon}{\partial S})_{M}
	&=\frac{3l^{2}\pi^{\frac{3}{2}}}{S^{\frac{5}{2}}}M-\frac{\pi l^{2}}{S^{2}}\notag\\
	&=\frac{\pi l^{2}-3(1+\varepsilon)S}{2S^{2}}.
\end{align}
The Hawking temperature at the event horizon is given by [6, 7],
\begin{align}
	T=\frac{1}{4\pi}(\frac{\partial f}{\partial r})_{r=r_{H}}.
\end{align}
We substitute the metric component (4) into the temperature (9) to obtain [20],
\begin{align}
	T=\frac{1}{4\pi}\frac{\pi l^{2}-3(1+\varepsilon)S}{l^{2}\sqrt{\pi S}}.
\end{align}
In virtue of the derivative (8) and the Hawking temperature (10), the combination is [20],
\begin{align}
	T(\frac{\partial S}{\partial\varepsilon})_{M}=\frac{S^{\frac{3}{2}}}{2\pi^{\frac{3}{2}}l^{2}}.
\end{align}
The Nariai limit means that $S\longrightarrow S_{N}$ if $T\longrightarrow 0$ and the limit leads the Eq.(10) to be [20],
\begin{align}
	S_{N}=\frac{\pi l^{2}}{3(1+\varepsilon)}.
\end{align}
Under the Nariai extremal condition, we substitute $S_{N}$ into the mass (6) [20],
\begin{align}
	M_{N}=\frac{l}{3\sqrt{3(1+\varepsilon)}}.
\end{align}
After proceeding the differentiation, the form is [20],
\begin{align}
	\frac{\partial M_{N}}{\partial\varepsilon}=-\frac{l}{6\sqrt{3}(1+\varepsilon)^{\frac{3}{2}}}.
\end{align}
The Nariai limit denoted as that $M\longrightarrow M_{N}$ while $S\longrightarrow S_{N}$ is imposed on the Eq.(11) [20],
\begin{align}
	\lim_{M\rightarrow M_{N}}(-T\frac{\partial S}{\partial\varepsilon})_{M}
	&=-\frac{S_{N}^{\frac{3}{2}}}{2\pi^{\frac{3}{2}}l^{2}}\notag\\
	&=-\frac{l}{6\sqrt{3}(1+\varepsilon)^{\frac{3}{2}}}.
\end{align}
By comparison between Eq.(14) and Eq.(15), the relation is obtained [20],
\begin{align}
	\frac{\partial M_{N}}{\partial\varepsilon}=\lim _{M\rightarrow M_{N}}(-T\frac{\partial S}{\partial\varepsilon})_{M},
\end{align}
which is a form of the Goon-Penco relation (1). It is significant that the universal relation is verified in the case of Schwarzschild-de Sitter black holes.

\section{The universal relation on the Schwarzschild-de Sitter black holes under GUP}

We are going to consider the universal relation on the Schwarzschild-de Sitter black holes within the frame of GUP. Within the microphysics regime, the GUP is chosen as [31, 35-39],
\begin{align}
\Delta x\Delta p\geq\frac{\hbar}{2}[1+(\frac{\beta l_{p}}{\hbar})^{2}\Delta p^{2}]
\end{align}
leading that,
\begin{align}
	y_{-}\leq y\leq y_{+}
\end{align}
where
\begin{align}
	y_{\pm}
	&=(\frac{l_{p}}{\hbar}\Delta p)_{\pm}\notag\\
	&=\frac{\Delta x}{\beta^{2}l_{p}}\pm\frac{\Delta\Delta x}{\beta^{2}l_{p}}\sqrt{1-(\frac{\beta l_{p}}{\Delta x})^{2}}.
\end{align}
Here $\beta$ is the positive dimensionless parameter modifying the Heisenberg uncertainty principle. The subscript $\pm$ corresponds to the sign $\pm$ in front of the second term in Eq.(18). The Planck length is $l_{p}=\sqrt{\frac{\hbar G}{c^{2}}}$ with velocity of light in the vacuum $c$. According to the generalized uncertainty principle (17), the corrected momentum difference can be obtained [26-39, 52, 53],
\begin{align}
	\Delta p'
	&=\frac{\hbar}{l_{p}}y_{-}\notag\\
	&=\frac{\hbar}{\beta^{2}l_{p}^{2}}\Delta x[1-\sqrt{1-(\frac{\beta l_{p}}{\Delta x})^{2}}].
\end{align}
Having combined the GUP (17) and the momentum difference (20), we estimate the black hole horizon $\Delta x'=2r'_{H}$ amended by the gravitation with the original horizon of black hole as the lower bound on the interval like $\Delta x=2r_{H}$ [36-39, 52, 53],
\begin{align}
	r'_{H}=r_{H}(1+\frac{\beta^{2}l_{p}^{2}}{16r_{H}^{2}}).
\end{align}
From the inequality (17), the appearance of $\beta$-term causes the Heisenberg uncertainty principle to be modified while the black hole horizon increases according to the expression (21). It should be pointed out that the distance difference $\Delta x'$ will be smaller if $\Delta p'=\frac{\hbar}{\ell_{p}}y_{+}$ is chosen and it is not reasonable to choose the $\Delta x'$ with $\Delta p'=\frac{\hbar}{\ell_{p}}y_{+}$ to constitute the black hole horizon like $\Delta x'=2r'_{H}$ [36-39].

It is fundamental to further the research on the thermal quantities of the Schwarzschild-de Sitter black hole under GUP. From Eq.(5) and Eq.(20), the GUP-corrected entropy of the black hole is,
\begin{align}
	S'=S+\frac{1}{8}\pi(\beta l_{p})^{2}+\frac{\pi^{2}(\beta l_{p})^{4}}{256S}.
\end{align}
The new horizon $r'_{H}$ is also thought as the root of $f(r'_{H})=0$ while the gravitational influence on the Heisenberg uncertainty can not be neglected, so the shifted mass of the black hole can also be written as,
\begin{align}
	M=\frac{\sqrt{S'}}{2\sqrt{\pi}}-\frac{1+\varepsilon}{2l^{2}}\frac{S'^{\frac{3}{2}}}{\pi^{\frac{3}{2}}}.
\end{align}
It is similar that the perturbative factor can changed as,
\begin{align}
	\varepsilon=-\frac{2l^{2}\pi^{\frac{3}{2}}}{S'^{\frac{3}{2}}}M+\frac{\pi l^{2}}{S'}-1.
\end{align}
Under the GUP, we perform the partial derivative to the perturbation (24) with respect to the entropy $S$ with the help of Eq. (22) while keep the mass $M$ fixed,
\begin{align}
	(\frac{\partial\varepsilon}{\partial S})_{M}=\frac{\pi l^{2}-3(1+\varepsilon)S'}{2S'^{2}}(1-\frac{\pi^{2}(\beta l_{p})^{4}}{256S^{2}}).
\end{align}
The Hawking temperature at the corrected horizon is [6, 7],
\begin{align}
	T'
	&=\frac{1}{4\pi}\frac{\partial f}{\partial r}|_{r=r'_{H}}\notag\\
	&=\frac{1}{4\pi}\frac{\pi l^{2}-3(1+\varepsilon)S'}{\sqrt{\pi S'}l^{2}}\notag\\
	&=\frac{1}{4\pi^{\frac{1}{2}}}\frac{S^{\frac{1}{2}}}{S+\frac{\pi(\beta l_{p})^{2}}{16}}-\frac{3(1+\varepsilon)}{4\pi^{\frac{3}{2}}l^{2}}\frac{S+\frac{\pi(\beta l_{p})^{2}}{16}}{S^{\frac{1}{2}}}.
\end{align}
It is clear that the corrected temperature $T'$ returns to be $T$ when $\beta=0$. Owing to the Nariai limit, $T'\longrightarrow 0$ then $S'\longrightarrow S'_{N}$, and the Eq.(26) tells us that,
\begin{align}
	S'_{N}=\frac{\pi l^{2}}{3(1+\varepsilon)}
\end{align}
meaning that $S'_{N}=S_{N}$ due to Eq.(12).

We arrange the temperature and the derivative subject to Eq.(24) and (25) respectively to obtain,
\begin{align}
	T'(\frac{\partial S}{\partial\varepsilon})_{M}=\frac{1}{2\pi^{\frac{3}{2}}l^{2}}\frac{S^{\frac{1}{2}}(S+\frac{\pi(\beta l_{p})^{2}}{16})^{2}}{S-\frac{\pi(\beta l_{p})^{2}}{16}}.
\end{align}
It is also obvious that the expression $T'(\frac{\partial S}{\partial\varepsilon})_{M'}$ will become $T(\frac{\partial S}{\partial\varepsilon})_{M}$ in Eq.(11) if $\beta=0$. We take the limit $M\longrightarrow M_{N}$ on $T'(\frac{\partial S}{\partial\varepsilon})_{M}$ and make use of Eq.(28) and Eq.(27) to find,
\begin{align}
	&\lim_{M'\rightarrow M_{N}}(-T'(\frac{\partial S}{\partial\varepsilon})_{M})\notag\\
	&=-\frac{1}{2\pi^{\frac{3}{2}}l^{2}}\frac{\frac{\pi l^{2}}{3(1+\varepsilon)}}{[\frac{\pi l^{2}}{3(1+\varepsilon)}-\frac{1}{4}\pi(\beta l_{p})^{2}]^{\frac{1}{2}}}\notag\\
	&\times\{(\frac{\pi l^{2}}{6(1+\varepsilon)}-\frac{1}{16}\pi(\beta l_{p})^{2})\notag\\
	&+[\frac{\pi l^{2}}{6(1+\varepsilon)}]^{\frac{1}{2}}[\frac{\pi l^{2}}{6(1+\varepsilon)}-\frac{1}{8}\pi(\beta l_{p})^{2}]^{\frac{1}{2}}\}.
\end{align}
According to the differentiation form (14) as the left hand of Goon-Penco relation, we rearrange the expression (29) as,
\begin{align}
	&\lim_{M'\rightarrow M_{N}}(-T'(\frac{\partial S}{\partial\varepsilon})_{M})\notag\\
	&=\frac{\partial M_{N}}{\partial\varepsilon}\notag\\
	&\times\sqrt{\frac{3(1+\varepsilon)}{\pi l^{2}}}[\frac{\pi l^{2}}{3(1+\varepsilon)}-\frac{1}{4}\pi(\beta l_{p})^{2}]^{-\frac{1}{2}}\notag\\
	&\times\{(\frac{\pi l^{2}}{6(1+\varepsilon)}-\frac{1}{16}\pi(\beta l_{p})^{2})\notag\\
	&+[\frac{\pi l^{2}}{6(1+\varepsilon)}]^{\frac{1}{2}}[\frac{\pi l^{2}}{6(1+\varepsilon)}-\frac{1}{8}\pi(\beta l_{p})^{2}]^{\frac{1}{2}}\}.
\end{align}
In order to keep the Goon-Penco universal relation, we have to look for the matching conditions. We choose the part of the limit (30) to be unit like,
\begin{align}
	\frac{\sqrt{3(1+\varepsilon)}}{\sqrt{\pi}l}[\frac{\pi l^{2}}{3(1+\varepsilon)}-\frac{1}{4}\pi(\beta l_{p})^{2}]^{-\frac{1}{2}}\notag\\
	\times\{\frac{\pi l^{2}}{6(1+\varepsilon)}-\frac{1}{16}\pi(\beta l_{p})^{2}\notag\\
	+[\frac{\pi l^{2}}{6(1+\varepsilon)}]^{\frac{1}{2}}[\frac{\pi l^{2}}{6(1+\varepsilon)}-\frac{1}{8}\pi(\beta l_{p})^{2}]^{\frac{1}{2}}\}=1,
\end{align}
then the equality $\frac{\partial M_{N}}{\partial\varepsilon}=\lim_{M\longrightarrow M_{N}}(-T\frac{\partial S}{\partial\varepsilon})_{M}=-\frac{1}{6\sqrt{3}(1+\varepsilon)^{\frac{3}{2}}}$ as a form of the Goon-Penco relation can be maintained in view of Eq. (14) and Eq. (30) [20]. We solve the necessary matching condition (31) to find $\beta l_{p}=0$, which means that the validity of the universal relation can only be kept under the original Heisenberg uncertainty. We compare Eq.(14) with Eq.(30) to show,
\begin{align}
	\frac{\partial M_{N}}{\partial\varepsilon}\neq\lim _{M\rightarrow M_{N}}(-T'(\frac{\partial S}{\partial\varepsilon})_{M}),
\end{align}
because $\beta\neq0$ cannot hold the condition (31). We show that the inequality sign will be replaced as the equality when $\beta=0$. We cannot choose the special thermodynamic quantities as matching conditions limiting the black holes and the conditions can remove the terms involving the coefficient $\beta$, so the Goon-Penco relation cannot be kept when $\beta\neq0$. The $\beta$-term as correction to the uncertainty principle violates the Goon-Penco universal relation.

\section{The universal relation on the Schwarzschild-de Sitter black holes under EUP}

Here we continue our discussing the Goon-Penco relation on the Schwarzschild-de Sitter black holes in the context of EUP. The EUP is shown as the position-uncertainty corrections to the Heisenberg inequality and we can select [30, 41-44],
\begin{align}
\Delta x\Delta p\geq\frac{\hbar}{2}(1+\frac{\alpha}{L_{*}^{2}}\Delta x^{2})
\end{align}
where $\alpha$ is a constant of order unit. It is interesting to discuss the extra term. The EUP affects the phase transitions of Schwarzschild and Reissner-Nordstrom black holes [42]. The bounds on the EUP scale can be investigated through astrophysical measurements [43]. In the process of exploring the EUP black holes with gravitational lensing, the fundamental scale of EUP from Sgr A* is limited as $~10^{10}m$ in the cases of deflection lensing [44]. There also exists a lower bound on the EUP scale as $~10^{13}m$ from the supermassive black holes with deflection lensing in the strong field limit [44] and $L_{*}$ is viewed as a large fundamental distance scale. The shift involving the ratio of distance difference and distance scale is extremely small. We solve the inequality (33) to introduce [30, 41-44],
\begin{align}
	\Delta x_{-}\leq\Delta x\leq\Delta x_{+}
\end{align}
where
\begin{align}
	\Delta x_{\pm}=\frac{L_{*}^{2}}{\hbar\alpha}\Delta p(1\pm\sqrt{1-\frac{\hbar^{2}\alpha}{L_{*}^{2}}\frac{1}{\Delta p^{2}}}).
\end{align}
The subscript $\pm$ refers to the sign $\pm$ in the righthand of the equation above. According to the scheme of Ref.[28], the $\Delta \overline{x}=\Delta x_{-}$ as a solution to the inequality (32) is selected while the momentum uncertainty $\Delta p$ is replaced, then the distance interval can be estimated as,
\begin{align}
	\Delta\overline{x}=\frac{\Delta x}{1+\frac{\alpha}{L_{*}^{2}}\Delta x^{2}}.
\end{align}
We also choose the original black hole horizon like $\Delta x=2r_{H}$ and the EUP-corrected horizon like $\Delta\overline{x}=2\overline{r}_{H}$ and express [30, 41-44],
\begin{align}
	\overline{r}_{H}=\frac{r_{H}}{1+\frac{4\alpha}{L_{*}^{2}}r_{H}^{2}}.
\end{align}
If the EUP effect disappears as $\alpha=0$, the corrected horizon $\overline{r}_{H}$ will be reverted to the original ones. The $\alpha$-term from the EUP shortens the black hole size according to Eq.(36). From Eq.(5) and (36), the entropy of black hole is changed as [30, 41-44],
\begin{align}
	\overline{S}=\frac{S}{(1+\frac{4\alpha}{\pi L_{*}^{2}}S)^{2}}.
\end{align}
We consider the equation $f(\overline{r}_{H})=0$ from the component function of metric to write the mass again [30, 41-44],
\begin{align}
	M=\frac{\sqrt{\overline{S}}}{2\sqrt{\pi}}-\frac{1+\varepsilon}{2l^{2}}\frac{\overline{S}^{\frac{3}{2}}}{\pi^{\frac{3}{2}}}
\end{align}
leading the perturbative parameter,
\begin{align}
	\varepsilon=\frac{2l^{2}\pi^{\frac{3}{2}}}{\overline{S}^{\frac{3}{2}}}M+\frac{\pi l^{2}}{\overline{S}}-1.
\end{align}
It is similar the partial derivative to the parameter (37) with respect to the entropy $S$ with the fixed mass $M$,
\begin{align}
	(\frac{\partial\varepsilon}{\partial S})_{M}&=\frac{1}{2}(1-\frac{4\alpha}{\pi L_{*}^{2}}S)\notag\\
	&\times[\frac{\pi l^{2}}{S^{2}}(1+\frac{4\alpha}{\pi L_{*}^{2}}S)-\frac{3(1+\varepsilon)}{S}(1+\frac{4\alpha}{\pi L_{*}^{2}}S)^{-1}]
\end{align}
under the EUP influence. It is necessary to consider the Hawking temperature with the EUP-corrected horizon,
\begin{align}
	\overline{T}
	&=\frac{1}{4\pi}\frac{\partial f}{\partial r}|_{r=\overline{r}_{H}}\notag\\
	&=\frac{1}{4\pi}\frac{\pi l^{2}-3(1+\varepsilon)\overline{S}}{\sqrt{\pi\overline{S}}l^{2}}\notag\\
	&=\frac{S^{\frac{3}{2}}}{4\pi^{\frac{3}{2}}l^{2}}[\frac{\pi l^{2}}{S^{2}}(1+\frac{4\alpha}{\pi L_{*}^{2}}S)-\frac{3(1+\varepsilon)}{S}(1+\frac{4\alpha}{\pi L_{*}^{2}}S)^{-1}].
\end{align}
When $\alpha=0$, the temperature (42) will become the same as Eq.(10). As a result of the Nariai limit, $\overline{S}\longrightarrow \overline{S}_{N}$ with $\overline{T}=0$, and the expression of $\overline{T}$ indicates that,
\begin{align}
	\overline{S}_{N}=\frac{\pi l^{2}}{3(1+\varepsilon)}
\end{align}
and $\overline{S}=\overline{S}_{N}$ because of Eq.(12).

It is useful to constitute the term $\overline{T}(\frac{\partial S}{\partial\varepsilon})_{M}$ with the help of Eq.(41) and Eq.(42),
\begin{align}
	\overline{T}(\frac{\partial S}{\partial\varepsilon})_{M}=\frac{1}{2\pi^{\frac{3}{2}}l^{2}}\frac{S^{\frac{3}{2}}}{1-\frac{4\alpha}{L_{*}^{2}}S}.
\end{align}
With $\alpha$-term vanishing, $\overline{T}(\frac{\partial S}{\partial\varepsilon})_{M}$ also returns to be $T(\frac{\partial S}{\partial\varepsilon})_{M}$ formulated by Eq.(11). With the limitation $M\longrightarrow M_{N}$ and the special entropy (42), it is shown that,
\begin{align}
	\lim_{M\rightarrow M_{N}}(-\overline{T}(\frac{\partial S}{\partial\varepsilon})_{M})
	&=-\frac{1}{2\pi^{\frac{3}{2}}l^{2}}\frac{1}{[\frac{3(1+\varepsilon)}{\pi l^{2}}-\frac{16\alpha}{\pi L_{*}^{2}}]^{\frac{1}{2}}}\notag\\
	&\times\{-\frac{\pi L_{*}^{2}}{4\alpha}+\frac{3(1+\varepsilon)}{2\pi l^{2}}(\frac{\pi L_{*}^{2}}{4\alpha})^{2}\notag\\
	&-\frac{3(1+\varepsilon)}{2\pi l^{2}}(\frac{\pi L_{*}^{2}}{4\alpha})^{2}\notag\\
	&\times[1-\frac{16\alpha l^{2}}{3(1+\varepsilon)L_{*}^{2}}]^{\frac{1}{2}}\}.
\end{align}
In view of the differentiation of mass under the Narial extremal condition with respect to the perturbation $\varepsilon$ like Eq. (14), we also rewrite the limitation (45) as,
\begin{align}
	\lim_{M\rightarrow M_{N}}(-\overline{T}(\frac{\partial S}{\partial\varepsilon})_{M})
	&=\frac{\partial M_{N}}{\partial\varepsilon}\notag\\
	&\times\frac{3\sqrt{3}(1+\varepsilon)^{\frac{3}{2}}}{\pi^{\frac{3}{2}}l^{3}}\frac{1}{[\frac{3(1+\varepsilon)}{\pi l^{2}}-\frac{16\alpha}{\pi L_{*}^{2}}]^{\frac{1}{2}}}\notag\\
	&\times\{-\frac{\pi L_{*}^{2}}{4\alpha}+\frac{3(1+\varepsilon)}{2\pi l^{2}}(\frac{\pi L_{*}^{2}}{4\alpha})^{2}\notag\\
	&-\frac{3(1+\varepsilon)}{2\pi l^{2}}(\frac{\pi L_{*}^{2}}{4\alpha})^{2}\notag\\
	&\times[1-\frac{16\alpha l^{2}}{3(1+\varepsilon)L_{*}^{2}}]^{\frac{1}{2}}\}.
\end{align}
The equality from the universal thermodynamic extremality relation (1) with Eq. (14) and (45) wants the part of right hand of Eq. (46) to be equal to the unit like,
\begin{align}
	\frac{3\sqrt{3}(1+\varepsilon)^{\frac{3}{2}}}{\pi^{\frac{3}{2}}l^{3}}\frac{1}{[\frac{3(1+\varepsilon)}{\pi l^{2}}-\frac{16\alpha}{\pi L_{*}^{2}}]^{\frac{1}{2}}}\notag\\
	\times\{-\frac{\pi L_{*}^{2}}{4\alpha}+\frac{3(1+\varepsilon)}{2\pi l^{2}}(\frac{\pi L_{*}^{2}}{4\alpha})^{2}\notag\\
	-\frac{3(1+\varepsilon)}{2\pi l^{2}}(\frac{\pi L_{*}^{2}}{4\alpha})^{2}\notag\\
	\times[1-\frac{16\alpha l^{2}}{3(1+\varepsilon)L_{*}^{2}}]^{\frac{1}{2}}\}=1
\end{align}
leading the equation combining perturbative parameter $\varepsilon$, cosmological radius $l$ and the EUP factor $\frac{\alpha}{L_{*}^{2}}$ as follow,
\begin{align}
	[\frac{\pi L_{*}^{2}}{4\alpha}-\frac{1}{8}\frac{\pi l^{2}}{3(1+\varepsilon)}]^{2}+\frac{63}{64}(\frac{\pi l^{2}}{3(1+\varepsilon)})^{2}=0.
\end{align}
The Eq.(48) as matching condition tells us that all real quantities like $\varepsilon$, $l$ and $\frac{\alpha}{L_{*}^{2}}$ can not support this equation. Within the frame of EUP, no matching condition can help to hold the universal relation. According to Eq.(46), the terms with $\alpha$ result in the violation of the universal relation,
\begin{align}
	\frac{\partial M_{N}}{\partial\varepsilon}\neq\lim_{M\rightarrow M_{N}}(-\overline{T}(\frac{\partial S}{\partial\varepsilon})_{M}).
\end{align}
It is obvious that the relation will recover if $\alpha=0$. The EUP breaks the universal relations. The generalization of Heisenberg uncertainty principle with momentum difference or distance interval does not hold the Goon-Penco relation.

\section{Conclusion}

We derive and compute the partial derivative to the mass, Hawking temperature and entropy of Schwarzschild-de Sitter black holes with respect to the perturbative parameter with Nariai limit under the corrected Heisenberg uncertainty principle to test the universal relation. We find that the Goon-Penco relation will not be tenable unless the influences from GUP or EUP are ignored. The combination of the generalizations of Heisenberg uncertainty principle with momentum difference or distance interval respectively and the black hole properties violates the Goon-Penco equation instead of becoming the matching conditions which maintains the universal relation. We argue that the corrected Heisenberg's uncertainty principles violate the validity of the universal relation. The similar investigations can be performed for the Reissner-Nordstrom-de Sitter and Kerr-de Sitter black holes with the same limit and the other black holes in future. It is to be expected that the additional terms involving the charge or angular momentum under GUP or EUP can not be cancelled with different value selection of the black hole constitutes such as mass, charge and angular momentum. The GUP- or EUP-corrected terms can also lead the Goon-Penco relation to be untenable.

\vspace{1cm}
\noindent \textbf{Acknowledge}

This work is partly supported by the Shanghai Key Laboratory of
Astrophysics 18DZ2271600.

\end{document}